\documentclass[10pt,journal,compsoc]{IEEEtran}
\usepackage{amsmath,amsfonts}
\usepackage{algorithm}
\usepackage[noend]{algpseudocode}
\usepackage{array}
\usepackage{xcolor}
\definecolor{darkred}{rgb}{0.7,0,0}
\definecolor{darkgreen}{rgb}{0,0.7,0}
\definecolor{darkblue}{rgb}{0,0,0.7}
\usepackage[colorlinks,bookmarks=false,linkcolor=darkred,citecolor=darkgreen,urlcolor=darkblue]{hyperref}
\usepackage{url}
\usepackage{graphicx}
\usepackage[square,numbers,sort&compress]{natbib}
\usepackage{booktabs}
\usepackage{paralist}
\DeclareGraphicsExtensions{.png,.jpg,.pdf}
\graphicspath{{figures/}{figures/confusion_matrices/}}
\newcolumntype{P}[1]{>{\raggedright\arraybackslash}p{#1}}

\begin{document}

\title{The Repeated-Stimulus Confound in Electroencephalography}

\author{Jack~A.~Kilgallen,
  Barak~A.~Pearlmutter,
  and~Jeffrey~Mark~Siskind,~\IEEEmembership{Senior~Member,~IEEE}%
  \thanks{J.~A.~Killgallen and B.~A.~Pearlmutter are with Hamilton
    Institute, Maynooth University, Maynooth, Ireland.
    B.~A.~Pearlmutter is also with Department of Computer Science,
    Maynooth University, Maynooth, Ireland.
    J.~M.~Siskind is with Elmore Family School of Electrical and
    Computer Engineering, Purdue University, West Lafayette, IN 47907-2035,
    USA.\\
    E-mail: \href{mailto:jack.eatonkilgallen.2019@mumail.ie}{jack.eatonkilgallen.2019@mumail.ie},
    \href{mailto:barak@cs.nuim.ie}{barak@cs.nuim.ie},
    \href{mailto:qobi@qobi.org}{qobi@qobi.org}}%
  \thanks{J.~M.~S. conceptualized the study with input from B.~A.~P.;
    J.~A.~K. devised and implemented the decoding analyses; J.~A.~K. performed
    all statistical analyses; B.~A.~P. and J.~M.~S. supervised the
    project; J.~A.~K. drafted the manuscript; all authors reviewed and
    edited the final version.}%
  \thanks{B.~A.~P. and J.~M.~S contributed equally to this work.}
  \thanks{This work has been submitted to the IEEE for possible publication.}}

\markboth{}%
{Kilgallen \etal: The Repeated-Stimulus Confound in Electroencephalography}

\IEEEpubid{}

\maketitle
\thispagestyle{empty}


\begin{abstract}
In neural-decoding studies, recordings of participants' responses to stimuli are used to train models. In recent years, there has been an explosion of publications detailing applications of innovations from deep-learning research to neural-decoding studies. The data-hungry models used in these experiments have resulted in a demand for increasingly large datasets. Consequently, in some studies, the same stimuli are presented multiple times to each participant to increase the number of trials available for use in model training. However, when a decoding model is trained and subsequently evaluated on responses to the same stimuli, stimulus identity becomes a confounder for accuracy. We term this the \emph{repeated-stimulus confound}. We identify a susceptible dataset, and 16 publications which report model performance based on evaluation procedures affected by the confound. We conducted experiments using models from the affected studies to investigate the likely extent to which results in the literature have been misreported. Our findings suggest that the decoding accuracies of these models were overestimated by between 4.46--7.42\%. Our analysis also indicates that per 1\% increase in accuracy under the confound, the magnitude of the overestimation increases by 0.26\%. The confound not only results in optimistic estimates of decoding performance, but undermines the validity of several claims made within the affected publications. We conducted further experiments to investigate the implications of the confound in alternative contexts. We found that the same methodology used within the affected studies could also be used to justify an array of pseudoscientific claims, such as the existence of extrasensory perception.
\end{abstract}

\maketitle


\section{Introduction}

\IEEEPARstart{R}{ecent} work \cite{kaneshiro_2015_representational} conducted
an original study of human perceptual and conceptual representation of object
classes.
As part of this study, a large dataset of electroencephalography (EEG)
recordings were taken from human subjects viewing image stimuli.
This dataset contained EEG responses of ten subjects, each undergoing
5,184 trials.
Each trial presented one of~72 images spanning~6 classes as a
stimulus (Fig.~\ref{fig:sud-category-pseudocategory-structure}).
Each of the~72 images was presented~72 times to each subject.
This study performed both 6-way (object class depicted in the stimulus image)
and 72-way (stimulus image identity) classification on the EEG recordings, using
this as a first step towards an analysis of representation similarity.

This study publicly released their dataset.
Since then, at least~18 additional papers \cite{bagchi_adequately_2021,
  bagchi_eeg-convtransformer_2022, luo_dual-branch_2023,
  kalafatovich_decoding_2020, kalafatovich_learning_2023,
  kalafatovich_subject-independent_2021, fares_brain-media_2020,
  jiang_brain-media_2021, bobe_single-trial_2018, xue_hybrid_2024,
  karimi-rouzbahani_temporal_2021, karimi-rouzbahani_when_2022,
  yavandhasani_visual_2022, zheng_evoked_2020, deng_eeg-based_2023,
  ahmadieh_hybrid_2023, ahmadieh_visual_2024, singh_learning_2024}
have been published using this dataset.
While the original study trained and tested classifiers primarily as a stepping
stone to representation similarly analysis (RSA), this subsequent work treated
stimulus classification from EEG recordings as the primary goal.
Most subsequent work solely studied 6-way classification of object
class depicted in the stimulus images.

The original study partitioned their dataset into a number of
training-set/test-set splits.
Each split contained a mix of EEG recordings elicited from all~72
stimulus images.
While this is not, strictly speaking, training on the test set---because the
training and test sets in each split contained disjoint sets of recordings---it
nonetheless introduced a confound because different EEG recordings from the
same subject viewing the same stimulus image are correlated, much in the same
fashion as data augmentation that is common when training computer-vision
models (e.g., slight variants of the same image), though in computer-vision the
augmented images are machine generated while here the variation is due to
brain processing.
We call this the \emph{repeated-stimulus confound} (RSC).
This confound leads to strongly overestimated classification accuracy.
All subsequent work appears to use these splits.

Of the~19 papers that use the dataset from the original study, including the
original study itself, 18~appear to use the dataset in a confounded fashion.
For one \cite{singh_learning_2024}, we were not able to determine
whether or not they use the dataset in a confounded fashion due to
insufficient detail.
Seven of these~19 papers \cite{kaneshiro_2015_representational,
  kalafatovich_decoding_2020, bagchi_adequately_2021,
  bagchi_eeg-convtransformer_2022, deng_eeg-based_2023,
  luo_dual-branch_2023, kalafatovich_learning_2023} either have code
or sufficiently detailed descriptions available to support replication;
the rest do not.
We repeat the experiments in all seven papers.
We find that:
\begin{compactenum}
\item Classification accuracy is severely inflated for all models
  (Table~\ref{tab:category-decoding-confounded-hp} and
  Fig.~\ref{fig:accuracy-bias-results}).
\item The severity of inflation for all models is statistically
  significant (Table~\ref{tab:bias-hypothesis-tests}).
\item The severity of the inflation increases substantially with the
  accuracy of the decoding model in a statistically significant
  fashion (Fig.~\ref{fig:bias-vs-confounded-accuracy} and
  Table~\ref{tab:bias-vs-confounded-lme}).
\item The classes become less separable in LDA, the model used by
  used in the original study, when the confound is removed
  (Fig.~\ref{fig:confounded-vs-unconfounded-confusion-matrix}).
  This calls their representation similarity analysis into question.
\item Per-class classification accuracy with the confound varies
  inversely compared to that without the confound.
  This indicates that the difficulty of decoding the hardest object
  categories has been substantially underestimated in the literature
  (Fig.~\ref{fig:category-bias-accuracy}).
\item The severity of the inflation is highly dependent on the
  distinction between classes, for most class pairs
  (Table~\ref{tab:category-bias-lme}).
\item In all models, the confound allows decoding artificial
  categories that don't exist
  (Tables~\ref{tab:pseudocategory-decoding-results}
  and~\ref{tab:pseudocategory-hypothesis-tests}).
  This allows using the confound to support potentially false claims.
\end{compactenum}
Collectively, all of the advances in classification accuracy of the~6 papers
that use the dataset from the original study, over the original study,
appear to vanish when the confound is removed.

Beyond methods from these seven papers, we repeat our analyses on six additional
commonly used machine-learning models and observe that the above seven issues
still occur.
This suggests all models, including those in the~12 papers for which we lack
code and sufficiently detailed descriptions to replicate, will suffer from the
RSC\@.

These~19 problematic papers were published across~18 different venues,
nine of which are in IEEE Xplore (Table~\ref{tab:venues}).
The severity of the issue, the fact that the issue arises from
machine learning applied to data derived from image stimuli, and the
need for prominent exposure to warn the community of the issue,
dictates publication in TPAMI.
Further, TPAMI and other IEEE computer-vision venues including CVPR, ICCV,
and the International Conference on Automatic Face and Gesture
Recognition, have published other work on analyzing EEG recordings
elicited from image stimuli \cite{spampinato2017, palazzo2017,
  palazzo2020a, palazzo2021, palazzo2024} which has been found
to be confounded for different reasons, as detailed by work published in CVPR and
TPAMI \cite{li2021, ahmed2021, ahmed2022, bharadwaj2023}.
As confounds have been found that invalidate dozens of published papers on EEG
decoding from image stimuli, the community needs to be made aware of the
issue.

\begin{table*}
  \centering
  \caption{Venues where affected studies have been published.
    Impact factors for journals from Impact factors are from JCR (2024).
    Ranks for conferences from Google Scholar Metrics (2025).}
  \label{tab:venues}
  \begin{scriptsize}
  \begin{tabular}{llrll}
    \toprule
    publication & model & IF & rank & venue\\
    \midrule
    \cite{ahmadieh_hybrid_2023} & & 4.5 & & Neural Computing and Applications\\
    \cite{ahmadieh_visual_2024} & & 4.9 & & Biomedical Signal Processing and Control\\
    \cite{bagchi_adequately_2021} & AW1DCNN & & & IEEE European Signal Processing Conference\\
    \cite{bagchi_eeg-convtransformer_2022} & EEGCT & 7.6 & & Pattern Recognition\\
    \cite{bobe_single-trial_2018} & & & & IEEE Engineering and Telecommunication\\
    \cite{deng_eeg-based_2023} & RLSTM & 4.9 & & Biomedical Signal Processing and Control\\
    \cite{fares_brain-media_2020} & & & 2 Multimedia & International Conference on Multimedia\\
    \cite{jiang_brain-media_2021} & & 9.7 & & IEEE Transactions on Multimedia\\
    \cite{kalafatovich_subject-independent_2021} & & & & IEEE International Winter Conference on Brain-Computer Interface\\
    \cite{kalafatovich_decoding_2020} & ADCNN & & 1 Automation \& Control Theory &
    IEEE International Conference on Systems, Man, and Cybernetics\\
    \cite{kalafatovich_learning_2023} & TSCNN & 5.2 & & IEEE Transactions on Neural Systems and Rehabilitation Engineering\\
    \cite{kaneshiro_2015_representational} & LDA & 2.6 & & PLOS ONE\\
    \cite{karimi-rouzbahani_when_2022} & & 3.2 & & Frontiers in Neuroscience\\
    \cite{karimi-rouzbahani_temporal_2021} & & 2.1 & & Neural Computation\\
    \cite{luo_dual-branch_2023} & STST & 9.9 & & IEEE Transactions on Industrial Informatics\\
    \cite{singh_learning_2024} & & & 6 Computer Vision \& Pattern Recognition &
    IEEE Winter Conference on Applications of Computer Vision\\
    \cite{xue_hybrid_2024} & & 3.9 & & Scientific Reports\\
    \cite{yavandhasani_visual_2022} & & 4.5 & & IEEE Transactions on Biomedical Engineering\\
    \cite{zheng_evoked_2020} & & & & International Conference on Neural Information Processing\\
    \bottomrule
  \end{tabular}
  \end{scriptsize}
\end{table*}

\section{Overview}

In a decoding experiment, if responses to the same stimuli are used to train and subsequently evaluate a decoding model, then those stimuli indirectly influence the model through the training data, and the evaluation of its accuracy through the test data. Consequently, stimulus identity is a confounder for decoding accuracy. We refer to this phenomenon as the \emph{repeated-stimulus confound} (RSC). While the RSC may affect the estimation of accuracy in almost any single-trial decoding task, in this study we limit our focus to the investigation of a specific instance of the confound which affects several publications on single-trial EEG object-category decoding.

Performing single-trial object-category decoding requires a stimulus set with multiple examples of objects belonging to each category of interest. We refer to these examples as \emph{exemplars} of the category. One approach to mitigating the impact of trial-to-trial variation on decoding analyses is to use a large stimulus set which features many exemplars of each category, and then record a single response to each exemplar. An advantage of this approach is that it promotes generalization to novel exemplars, as models are trained on a wide variety of examples. However, it can be both time-consuming and challenging to curate a large stimulus set which lacks spurious correlations between the category label and other variables such as contrast, brightness, size, or other features which are not representative of the category. A less time-and-resource intensive alternative is to use a small stimulus set, and then record multiple responses to each exemplar to improve the signal-to-noise ratio of the data. However, while this approach may be effective at reducing the impact of trial-to-trial variation, it introduces a new challenge.

\begin{figure*}
  \centering
  \includegraphics[width=0.75\textwidth, trim={2cm, 2cm, 2cm, 2cm}, clip]{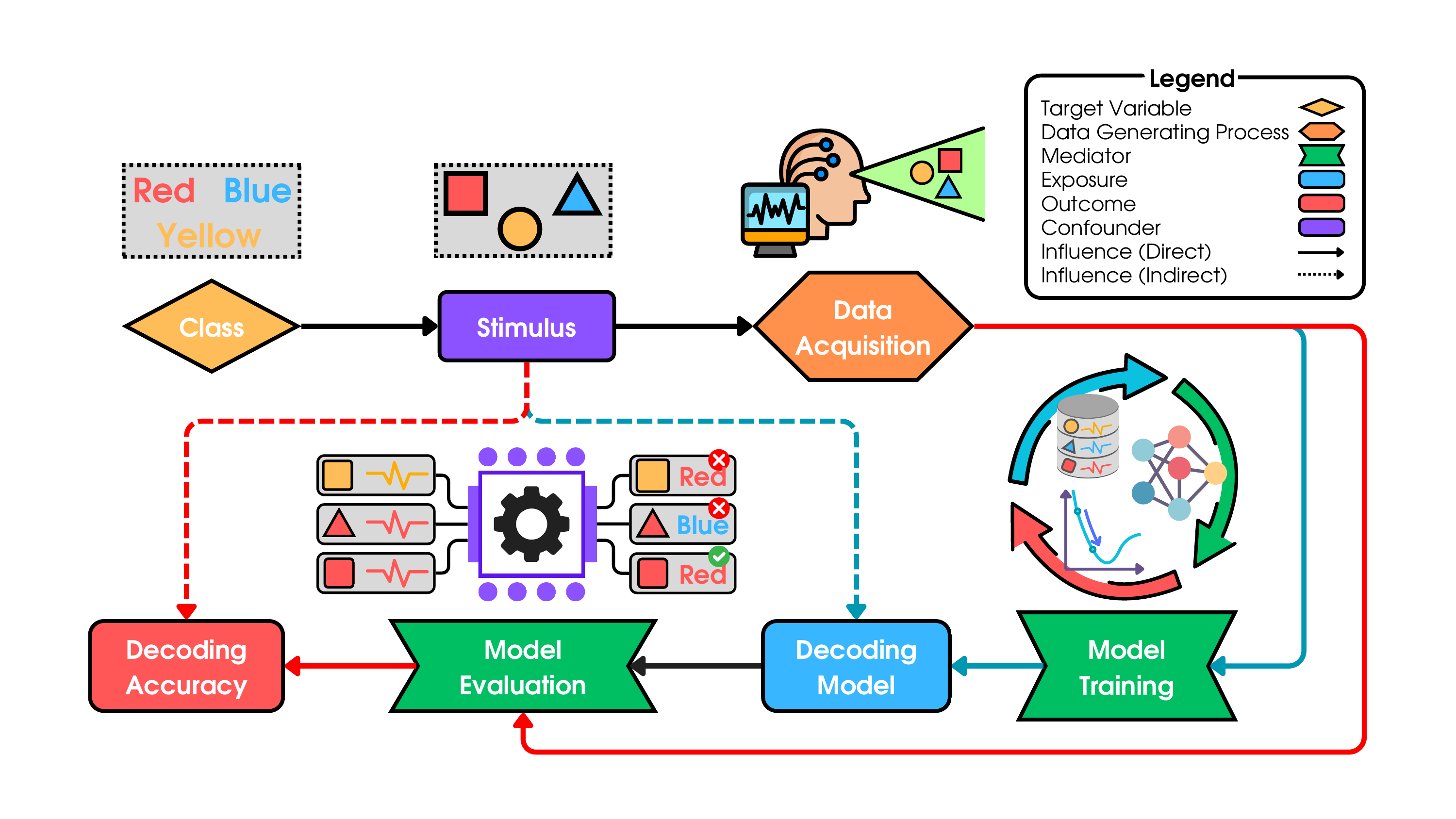}
  \caption{\textbf{Causal diagram of the RSC.} Suppose you wish to perform EEG object-color decoding. To conduct your experiment, you would need to record neural responses to stimuli of a predetermined set of colors. During the decoding analysis, you would then train a model using a subset of the data, and evaluate its accuracy using the remainder. However, suppose you have only a few stimuli of each color, but record multiple responses to each of them. Suppose also that you train, and subsequently evaluate, the model on responses to the same stimuli. Then, if the model correctly decodes a response to a given stimulus, like a red square, how would you know if it was decoding color, shape, or some combination of the two? Would it be as likely to correctly decode a response to a novel stimulus, like a red triangle? Unfortunately, as the causal diagram illustrates, since the stimuli influence both the model (through the blue path) and its accuracy (through the red path), the relationship between the two is confounded. Consequently, the representations learned by the model have not been properly validated, and its ability to decode responses to novel stimuli is unknown.}
  \label{fig:repeated-stimulus-confound-chart}
\end{figure*}

By way of analogy, consider a computer-vision classification model. Suppose we take a small dataset of images, and generate a larger dataset by creating noisy copies of each image. Since noisy copies of the same image are still highly correlated, they form dense, localized clusters in the feature space. Therefore, if we train a model using this dataset, the loss gradients will also be highly correlated, and the model will learn to overfit to features of specific images. If we evaluate the model on additional noisy copies of the same images, then our estimate of its performance will be, at least partially, based on the image-specific features it has learned. This is the essence of the repeated-stimulus confound. In recording a neural response, we are effectively applying a noisy transformation to the stimulus which generated it. And while the transformation may be  noisy, non-linear, and high-dimensional, transformations of the same stimulus may still form dense, localized clusters in the feature space. Therefore, if a category-decoding model is trained on a dataset with multiple responses per stimulus, it may learn stimulus-specific features of the neural responses, rather than features which generalize to the category as a whole. And, as Fig.~\ref{fig:repeated-stimulus-confound-chart} illustrates, if the model is evaluated using responses to the same stimuli, then the estimate of its decoding accuracy will be confounded by stimulus identity, and is not a reliable indicator of its ability to learn generalizable features of object categories.

\subsection{Publications affected by the repeated-stimulus confound}

The potential for the repetition of stimuli to confound decoding accuracy came to our attention when several publications which used the same EEG object-category decoding dataset were found to train and subsequently evaluate their models on responses to the same stimuli. This dataset, commonly referred to as the Stanford University dataset (SUD), contains EEG recordings of neural responses to images of natural objects, each of which is an example of a predefined category. The stimulus set was curated for use in an MEG study \cite{kriegeskorte_matching_2008}, and subsequently used to create the SUD for a study investigating the dynamics of visual object processing using representational similarity analysis (RSA) \cite{kaneshiro_2015_representational}.

To investigate if the confounded publications were isolated cases, or part of a more widespread issue, we conducted a comprehensive review of all publications which use the dataset to perform object-category decoding. We considered any decoding analysis which involved training, and subsequently evaluating, a model on responses to the same stimuli as susceptible to the RSC. Our audit revealed that 18 of the 19 papers which we identified, including the study for which the SUD was collected, performed confounded analyses \cite{kaneshiro_2015_representational,bagchi_adequately_2021,bagchi_eeg-convtransformer_2022,luo_dual-branch_2023,kalafatovich_decoding_2020,kalafatovich_learning_2023,kalafatovich_subject-independent_2021,fares_brain-media_2020, jiang_brain-media_2021, bobe_single-trial_2018,
xue_hybrid_2024,karimi-rouzbahani_temporal_2021,karimi-rouzbahani_when_2022,yavandhasani_visual_2022,zheng_evoked_2020,deng_eeg-based_2023,ahmadieh_hybrid_2023, ahmadieh_visual_2024}. The remaining work does not reference a specific evaluation procedure, however, it does report its decoding accuracy in relation to one of the publications we found to use a confounded decoding analysis \cite{singh_learning_2024}. The majority of these studies assess accuracy using a random, category-stratified, or unspecified 10-fold cross-validation method \cite{
  kaneshiro_2015_representational,bagchi_adequately_2021,bagchi_eeg-convtransformer_2022,luo_dual-branch_2023,yavandhasani_visual_2022,kalafatovich_decoding_2020,kalafatovich_learning_2023,kalafatovich_subject-independent_2021,fares_brain-media_2020, jiang_brain-media_2021,bobe_single-trial_2018,xue_hybrid_2024,karimi-rouzbahani_temporal_2021,karimi-rouzbahani_when_2022}. One other study performs 5-fold cross-validation \cite{yavandhasani_visual_2022}, and two more use training-test splits \cite{zheng_evoked_2020,deng_eeg-based_2023}. We also noted two publications which report decoding accuracy based on a test set with a single response to each stimulus, which would result in an estimate of accuracy which is unstable, due to trial-to-trial variation, and maximally confounded, as the model is trained on the highest possible number of responses per stimulus \cite{ahmadieh_hybrid_2023,ahmadieh_visual_2024}. Two publications, both by the same authors, identify the potential for the repetition of stimuli to bias the representations learned by models, and consequently restrict their analyses to a subset of the dataset \cite{karimi-rouzbahani_temporal_2021,karimi-rouzbahani_when_2022}. The RSC likely affects these papers substantially less than other work, as the levels of stimulus repetition are relatively low (twelve repetitions per stimulus), and additional datasets were used to validate their results. Similarly, one publication, which used the SUD in a novel decoding task, also identified the potential for the confound to bias category-decoding analyses, but correctly determined that it was not relevant to their study \cite{mccartney_zero-shot_2022}.

While the original study was published almost a decade ago, the subsequent papers were published within the last five years, with only one exception. We attribute this to the ongoing surge in interest in EEG-decoding via deep-learning, as the majority of these publications propose novel deep-learning architectures. Moreover, while the decoding model in the original study was only of relevance insofar as it facilitated further analyses, later publications are predominantly concerned with demonstrating the efficacy of novel deep-learning architectures, preprocessing techniques, or feature selection methods.

The approaches described in these publications comprise a broad range of innovative developments in EEG decoding. Moreover, the feature engineering techniques they employ demonstrate a rapid evolution in the field. The earliest models used the same temporal features as the original study \cite{kalafatovich_decoding_2020,bagchi_adequately_2021}, while successive models incorporate spatio-temporal features \cite{bagchi_eeg-convtransformer_2022}, and spectral features \cite{luo_dual-branch_2023}. One later approach integrates spatio-temporal dynamics by constructing graph representations in which nodes, representing electrodes, are connected if they satisfy functional connectivity criteria. The adoption of novel deep-learning architectures has been similarly expeditious. While early papers used models with simple 1D convolutions \cite{bagchi_adequately_2021}, later publications describe models which incorporate recurrent layers \cite{bagchi_adequately_2021}, attention mechanisms \cite{deng_eeg-based_2023}, and transformer networks \cite{luo_dual-branch_2023}.

However, given the susceptibility of these studies to the repeated-stimulus confound, the efficacy of the developments they present are called into question. Importantly, for many of the studies, the decoding accuracy they report is the primary metric by which the significance of their work is measured. Therefore, the confound undermines the validity of the claims made in these publications. And, while the uncertainty in both the absolute and relative performance of models from the affected publications is the most immediate consequence of the confound, it may not be the most salient. Our findings show that the accuracies of the most proficient models are those most overestimated under a confounded evaluation procedure. Our analyses also show that the relative difficulty of decoding different categories is substantially underestimated when the confound is present. Given these limitations, the claims about the potential utility of EEG-decoding models in real-world applications are questionable. In isolation, this may not be a significant issue, as no publication reports a decoding accuracy which is sufficiently high to be considered a breakthrough. However, in aggregate, they severely misrepresent the limitations of the deep-learning techniques they describe, the feature engineering techniques they employ, and potentially the utility of EEG as a decoding modality.

It is important to note that the RSC is not only of relevance to publications which use the SUD. Several other frequently used EEG datasets, such as THINGS-EEG \cite{grootswagers2022human}, THINGS-EEG2 \cite{gifford_large_2022}, and Alljoined1 \cite{xu_all_joined_2024}, also feature multiple responses per stimulus, and are consequently also susceptible to the RSC. Moreover, while the RSC as we describe it relates only to decoding analyses, it is likely the repetition of stimuli also affects analyses such as image reconstruction.

\subsection{A practical and RSC-robust approach to decoding analysis}

The factor which gives rise to the RSC is not the repetition of stimuli, but the reuse of stimuli to evaluate a model when they have been previously used to train it. Therefore, it is possible to mitigate the confound by ensuring that the model is trained and evaluated on disjoint stimuli. However, there are additional factors which must be considered to facilitate a robust evaluation procedure. To ensure that the accuracy of a model is not biased by the specific stimuli used in a test set, it is preferable to ensure each stimulus is used in a test set exactly once. Additionally, it is desirable that the training and test sets have the same distribution of class labels to prevent class imbalances from biasing model predictions. These considerations recommend the use of a class-stratified stimulus-grouped $k$-fold cross-validation scheme. This ensures that, over all folds, each stimulus is used to evaluate the model once. While similar procedures have been used extensively within the statistical literature, it is not commonly employed in decoding analyses. A necessary limitation of any solution to the RSC is that models cannot be trained on the full set of stimuli collected. Depending on the proportion of stimuli used in training, this may result in a significant reduction in the diversity of stimuli and may affect the generalizability of the representations learned by models. To address this issue, where feasible, we suggest the use of a leave-one-stimulus-per-class-out cross-validation strategy. This approach ensures models are trained on the maximum number of stimuli in each cross-validation fold, while still ensuring that the confound is mitigated.

\section{Materials and Methods}

In designing our decoding experiments, we had three primary aims. First, to quantify the likely extent to which the performance of models has been misreported within the affected literature. Second, to determine if stimulus repetition biases the representations learned by decoding models. And lastly, to investigate the broader implications of the continued use of methodologies susceptible to the RSC.

We performed our experiments using the SUD and a selection of models from the affected publications. We also included several additional models to reflect the range of approaches commonly featured within the broader EEG-decoding literature. To assess the bias in decoding accuracy due to the RSC, while mitigating the variability of the model training process, we devised a novel \emph{paired cross-validation} procedure. In each cross-validation fold, a single model is trained on a training set, and subsequently tested on two distinct test sets. One test set contains responses to the same stimuli which the model was trained on, and another which features responses to novel stimuli. Therefore, for each model, we gain both a confounded, and an unconfounded estimate of its decoding accuracy, and the discrepancy between the two is our estimate of the bias due to the RSC.

In addition to performing our object-category decoding experiments, we also devised a novel decoding task to investigate the implications of the confound in alternative contexts. In this task, we assigned a new category label to each stimulus such that the new object-categories lacked any meaningful distinction between categories. We refer to these new category labels as \emph{pseudocategories}. In our pseudocategory decoding experiments, we used the same paired cross-validation method as in our object-category decoding experiments. This allowed us to investigate the implications of the RSC in decoding analyses which attempt to decode non-existent properties of stimuli.

\subsection{The Stanford University dataset}

\begin{figure*}
  \hspace*{\fill}
  \begin{minipage}[t]{5.7cm}
    \centering
    \includegraphics[width=\columnwidth, trim={16cm, 0cm, 15cm, 0cm}, clip]{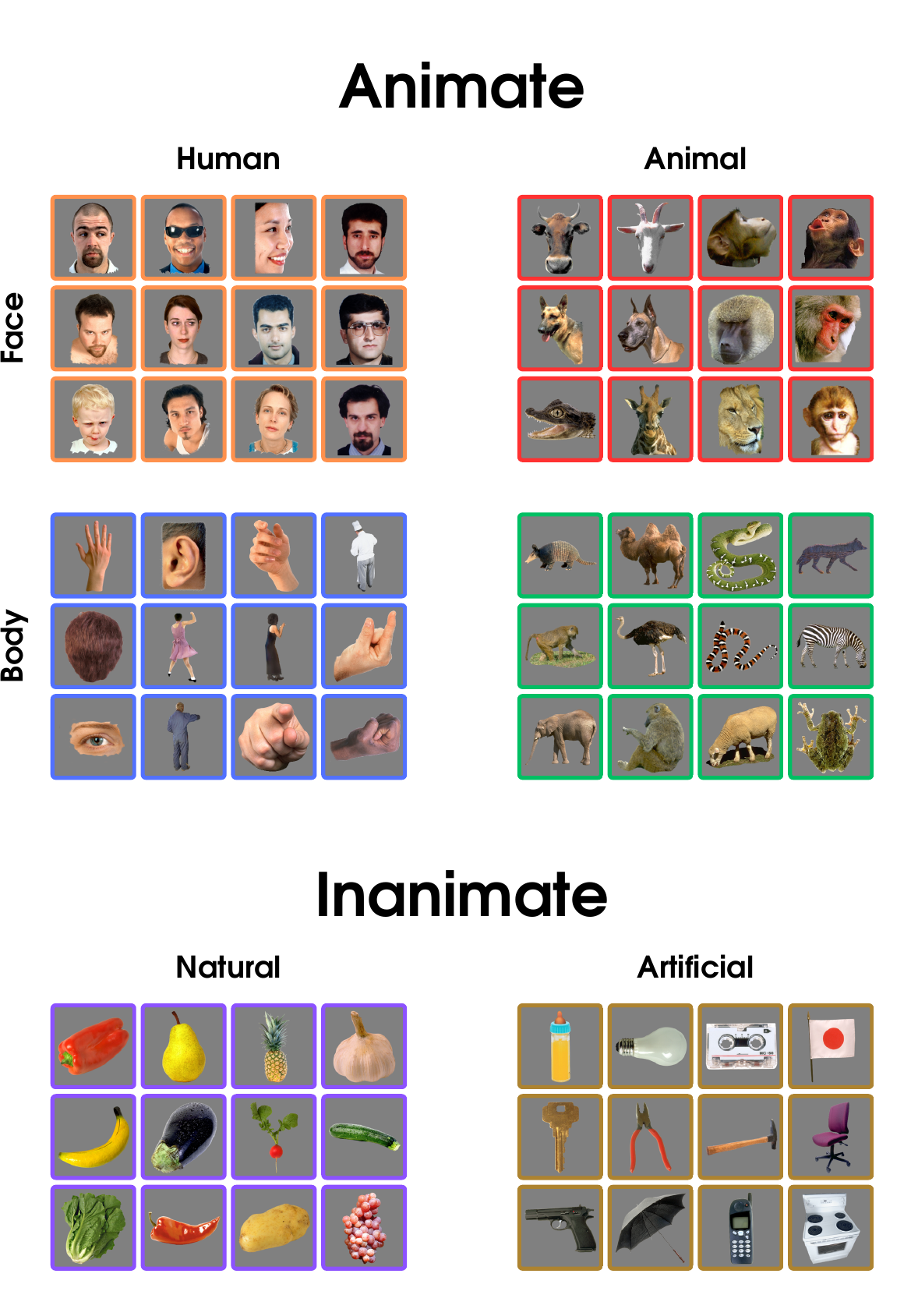}
  \end{minipage}
  \hspace*{\fill}
  \begin{minipage}[t]{5.7cm}
    \centering
    \includegraphics[width=\columnwidth, trim={16cm, 0cm, 15cm, 0cm}, clip]{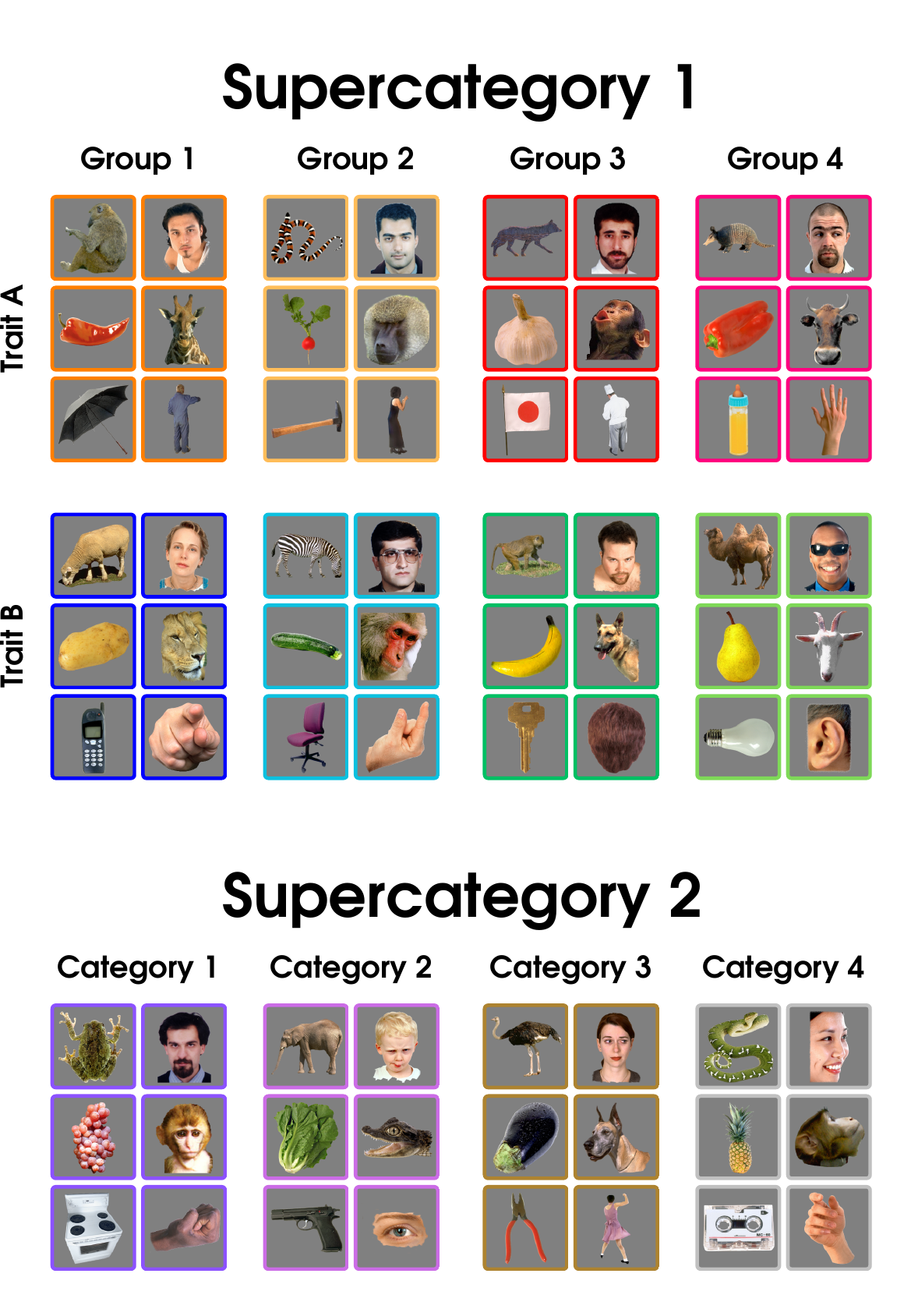}
  \end{minipage}
  \hspace*{\fill}
  \caption{\textbf{The category and pseudocategory structures of the SUD stimulus set.} The stimulus set used in the SUD is composed of 72 images of natural objects. In the original category structure of the SUD (left), these stimuli are evenly distributed across six categories: Human Body (HB), Human Face (HF), Animal Body (AB), Animal Face (AF), Fruit/Vegetable (FV), and Inanimate Object (IO). These can be further grouped into Animate and Inanimate supercategories. To create the pseudocategory structure (right), one stimulus from each of the original categories was assigned to form each new pseudocategory. This was done to create arbitrary category labels, which lack any inherent meaning or distinction. No two stimuli from the same category were assigned to the same pseudocategory, which resulted in twelve pseudocategories, each with six stimuli. This was done to prevent random variations in intraclass similarity from inducing legitimate distinctions between pseudocategories. Consequently, the stimuli within a pseudocategory are no more similar to each other than would be expected by random chance. In representations of category and pseudocategory structure, the colored frames delineate category, and were not presented to the subjects.}\vspace{-1em}
  \label{fig:sud-category-pseudocategory-structure}
\end{figure*}

The Stanford University Dataset (SUD) consists of preprocessed EEG recordings from 10 subjects, obtained while they viewed 72 images evenly distributed across 6 categories. To reduce the impact of trial-to-trial variation, each stimulus was presented 72 times to each subject, for a total of 5,184 trials per subject. The stimuli were recorded in six blocks of 864 trials each, over two sessions. In each block, all stimuli appeared 12 times, in a randomized order. Each stimulus was presented for 500~ms, followed by a 750~ms inter-stimulus interval, during which a gray background was displayed. The data was recorded using a 128-channel EEG system with a sampling rate of 1~kHz. The EEG signals were then preprocessed using a high-pass fourth-order Butterworth filter to attenuate frequencies below 1~Hz, and a low-pass Chebyshev Type I filter to attenuate frequencies above 25~Hz. The data was then subsampled to 62.5~Hz to reduce the computational cost of the analysis. Ocular artifacts were removed using the Bell-Sejnowski Infomax independent-component-analysis algorithm \cite{bell-sejnowski-1995}.  The 4 channels used to detect ocular artifacts were removed, and the remaining channels were converted to average reference. Finally, epochs of 496~ms post-stimulus response, time-locked to the onset of the stimulus, were extracted from the data. As a result, each trial is represented as a 124$\times$32 feature matrix, where the first dimension represents the 124 channels which were retained, and the second dimension represents the 32 time points of post-stimulus response for each trial. As mentioned previously, to perform our pseudocategory-decoding experiments, we created a new pseudocategorical structure for the stimulus set. By construction, as Fig.~\ref{fig:sud-category-pseudocategory-structure} demonstrates, the pseudocategories are devoid of any meaningful distinctions.

\subsection{Decoding models}

To perform the decoding experiments, a sample of models from affected publications were selected. The criterion for selection was that the model must originate from a work which featured single-trial subject-dependent object-category decoding on the SUD as the primary analysis. This decision was made to prioritize the inclusion of models from works for which the RSC is most relevant, while facilitating a consistent comparison of model performance in the context of the results reported in the affected publications. However, this selection was also limited by the availability of source code or, in the absence thereof, reasonably detailed descriptions of model architecture. The models selected for use in the experiments, as well as a brief description of their architecture, and any feature engineering methods they require, are detailed in Table~\ref{tab:affected-studies-models}. Additionally, several models commonly used in the wider EEG-decoding literature were also included for use in the experiments. These models were selected to investigate if more general EEG-decoding models are also susceptible to the repeated-stimulus confound. These included classical machine-learning models such as multi-class logistic regression (LR), k-Nearest Neighbors (kNN), and Support-Vector Classification (SVC). Also included were popular deep-learning models such as ShallowConvNet \cite{schirrmeister_deep_2017}, DeepConvNet \cite{schirrmeister_deep_2017} and EEGNet \cite{lawhern_eegnet_2018}.

\begin{table}
  \centering
  \caption{Models from affected studies included in our experiments.}
  \label{tab:affected-studies-models}
  \renewcommand{\arraystretch}{1.5}
  \begin{tabular}{@{\quad}l P{0.475\linewidth} c}
    \toprule
    \multicolumn{1}{l}{\textbf{Model}} &
    \textbf{Description} &
    \textbf{Accuracy}\textsuperscript{*}\\
    \midrule
    LDA \mbox{\cite{kaneshiro_2015_representational}} & A Linear Discriminant Analysis model trained using a one-vs-rest approach. The dimensionality of the data was reduced by applying principal component analysis (PCA). & 40.68\%\textsuperscript{\dag}\\
    ADCNN \mbox{\cite{kalafatovich_decoding_2020}} & A convolutional neural network (CNN), featuring two identical blocks of convolutional layers. One block is applied to data from all channels, while a parallel block is applied only to data from occipital channels. &  50.37\%\textsuperscript{\dag}\\
    AW1DCNN \mbox{\cite{bagchi_adequately_2021}} & A shallow CNN which uses 1-dimensional convolutional layers with a large quantity of filters to extract temporal features from the data. Additionally, the model features residual connections between blocks of convolutional layers. &  51.29\%\textsuperscript{\dag}\\
    EEGCT-Slim \mbox{\cite{bagchi_eeg-convtransformer_2022}} & A convolutional transformer featuring multi-head self attention. The data was spatially projected using an azimuthal equidistant projection, and applying Clough-Tocher interpolation to generate image-like activity maps for each time point. & 51.96\%\textsuperscript{\dag}\\
    EEGCT-Wide \mbox{\cite{bagchi_eeg-convtransformer_2022}} & A wide variant of the EEGCT-Slim architecture featuring convolutional layers, an increased number of filters, and additional self-attention heads. & 52.33\%\textsuperscript{\dag}\\
    RLSTM \mbox{\cite{deng_eeg-based_2023}} & A long short-term memory (LSTM) network which reuses LSTM layers by applying each layer to all channels. & 52.69\%\textsuperscript{\ddag}\\
    STST \mbox{\cite{luo_dual-branch_2023}} & A transformer network which uses self attention to fuse spatio-temporal and spectral features of the data. The spectral features of each channel were extracted by applying a continuous wavelet transform with a Morlet wavelet. & 54.82\%\textsuperscript{\dag}\\
    TSCNN \mbox{\cite{kalafatovich_learning_2023}} & A graph convolutional neural network which models the spatial stream of visual processing. Graph representations of the data were constructed by connecting channels with edges if they satisfied a spatial or functional connectivity criterion. & 54.28\%\textsuperscript{\dag}\\
    \bottomrule%
    \multicolumn{3}{p{0.85\linewidth}}{\small `*' The reported category-decoding accuracy on the SUD.}\\
    \multicolumn{3}{p{0.85\linewidth}}{\small `\dag' A 10-fold cross-validation procedure was used.}\\
    \multicolumn{3}{p{0.85\linewidth}}{\small `\ddag' A 90/10 train-test split was used.}
  \end{tabular}
\end{table}

\subsection{Training and evaluation procedures}

As mentioned previously, to obtain estimates of the bias imparted by the RSC, we devised a paired cross-validation procedure. The advantage of this method is that it provides both a confounded and unconfounded estimate of decoding accuracy for each trained model. During our analyses, we observed that our method of constructing paired cross-validation folds has utility in two key applications beyond our investigation. First, it serves as a template for investigating other confounds involving discrete hierarchical variables. Second, it provides a straightforward and stable approach for validating the stimulus-specific bias affecting decoding models. Therefore, in addition to the pseudocode in Algorithm~\ref{alg:paired-cv} we provide a full implementation in the supplementary materials.\footnote{\url{https://github.com/qobi/tpami2025}}

\begin{algorithm}
  \caption{Construction of paired cross-validation folds.}
  \label{alg:paired-cv}
  \begin{algorithmic}
    \State \textbf{Input:} $N_{\textbf{stim/cat}}$ leave-one-stimulus-per-category-out folds $\mathbf{idx}^{\alpha}$, and an equal number of stimulus-stratified folds $\mathbf{idx}^{\beta}$, where $N_{\textbf{stim/cat}}$ is the number of stimuli per category.
    \State \textbf{Output:}  $N_{\textbf{stim/cat}}$ paired cross-validation folds $\mathbf{idx}^{\gamma}$, the training set, unconfounded test set, and confounded test set of fold $i$ are indexed by $\mathbf{idx}^{\gamma}_{i, \lambda}$, $\mathbf{idx}^{\gamma}_{i, \alpha}$, and $\mathbf{idx}^{\gamma}_{i, \beta}$, respectively.
    \vspace{\baselineskip}
    \For{$i \in \{1, \ldots, N_{\textbf{stim/cat}}\}$}
      \State $\mathbf{idx}^{\gamma}_{i, \alpha} \gets \mathbf{idx}^{\alpha}_{i} \setminus \mathbf{idx}^{\beta}_{i}$
      \State $\mathbf{idx}^{\gamma}_{i, \beta} \gets \mathbf{idx}^{\beta}_{i} \setminus \mathbf{idx}^{\alpha}_{i}$
      \For{$i \in \{1, \ldots, N_{\textbf{stim/cat}}\}$}
        \If{$j \neq i$}
          \State $\mathbf{idx}^{\gamma}_{j, \lambda} \gets \mathbf{idx}^{\gamma}_{j, \lambda} \cup \mathbf{idx}^{\alpha}_{i}$
        \EndIf
      \EndFor
    \EndFor
  \end{algorithmic}
\end{algorithm}

In all of our experiments, hyperparameter tuning was performed using nested cross-validation. For each of the $k$ outer folds, the training set was split into $k-1$ stimulus-stratified inner folds. For each subject and outer fold of all experiments, a search was performed to find the hyperparameter values which minimized average validation loss over all nested folds. When available, the ranges of hyperparameters specified for each model in its corresponding publication were used during hyperparameter tuning. Similarly, where possible, any fixed hyperparameters reported were used. In the absence of adequate documentation, we used predetermined default values and ranges. Following hyperparameter tuning, a final model was trained using the combined training and validation sets, and then evaluated on both the confounded and unconfounded test sets.

\section{Results}

Tables~\ref{tab:category-decoding-confounded-hp} and \ref{tab:pseudocategory-decoding-results} present the results of our category and pseudocategory decoding experiments respectively. By convention, we report the decoding performance of each model in terms of its accuracy, and the standard deviation in accuracy across subjects. For each model in each experiment, mean confounded and unconfounded accuracy estimates are reported side by side for ease of comparison. Similarly, to quantify the bias imparted by the repeated-stimulus confound, we also report the mean difference between confounded and unconfounded estimates of accuracy, and the standard deviation of this difference across subjects.

\begin{table}
  \centering
  \caption{Confounded accuracy (CA), unconfounded accuracy (UA), and bias in the object-category-decoding experiments.}
  \label{tab:category-decoding-confounded-hp}
  \begin{tabular}{lccc}
\toprule
\textbf{Model} & \textbf{CA} & \textbf{UA} & \textbf{Bias} \\
\midrule
\textbf{Primary models} & & & \\
\quad LDA & 39.82 $\pm$ 6.94 & 35.36 $\pm$ 5.40 & 4.46 $\pm$ 1.98 \\
\quad ADCNN & 51.26 $\pm$ 8.50 & 43.32 $\pm$ 5.43 & 7.94 $\pm$ 3.48 \\
\quad AW1DCNN & 48.99 $\pm$ 7.01 & 42.34 $\pm$ 5.06 & 6.65 $\pm$ 2.45 \\
\quad EEGCT-Slim & 50.24 $\pm$ 7.55 & 42.82 $\pm$ 5.24 & 7.42 $\pm$ 2.90 \\
\quad EEGCT-Wide & 45.59 $\pm$ 6.41 & 39.76 $\pm$ 5.04 & 5.82 $\pm$ 2.10 \\
\quad RLSTM & 46.48 $\pm$ 6.86 & 40.43 $\pm$ 5.17 & 6.04 $\pm$ 2.51 \\
\quad STST & 45.19 $\pm$ 5.92 & 40.02 $\pm$ 4.60 & 5.17 $\pm$ 1.99 \\
\quad TSCNN & 45.42 $\pm$ 5.99 & 39.84 $\pm$ 4.62 & 5.58 $\pm$ 2.01 \\
\textbf{Additional models} & & & \\
\quad LR & 43.84 $\pm$ 5.95 & 38.12 $\pm$ 4.50 & 5.72 $\pm$ 1.92 \\
\quad kNN & 25.51 $\pm$ 2.07 & 24.07 $\pm$ 1.94 & 1.44 $\pm$ 0.84 \\
\quad SVC & 41.97 $\pm$ 5.29 & 36.56 $\pm$ 4.01 & 5.42 $\pm$ 2.04 \\
\quad ShallowConvNet & 37.73 $\pm$ 4.32 & 34.74 $\pm$ 3.59 & 2.99 $\pm$ 1.75 \\
\quad DeepConvNet & 43.47 $\pm$ 6.22 & 38.89 $\pm$ 5.25 & 4.58 $\pm$ 1.57 \\
\quad EEGNet & 49.17 $\pm$ 7.48 & 42.85 $\pm$ 5.34 & 6.32 $\pm$ 2.69 \\
\bottomrule
\end{tabular}

\end{table}

\begin{table}
  \centering
  \caption{Confounded accuracy (CA), unconfounded accuracy (UA) and bias in the pseudocategory-decoding experiments.}
  \label{tab:pseudocategory-decoding-results}
  \begin{tabular}{lccc}
\toprule
\textbf{Model} & \textbf{CA} & \textbf{UA} & \textbf{Bias} \\
\midrule
\textbf{Primary models} & & & \\
\quad LDA & 10.90 $\pm$ 1.68 & 8.79 $\pm$ 0.59 & 2.11 $\pm$ 2.01 \\
\quad ADCNN & 19.50 $\pm$ 4.79 & 8.49 $\pm$ 0.79 & 11.01 $\pm$ 4.86 \\
\quad AW1DCNN & 15.32 $\pm$ 3.29 & 8.40 $\pm$ 0.69 & 6.93 $\pm$ 3.51 \\
\quad EEGCT-Slim & 16.58 $\pm$ 3.85 & 8.17 $\pm$ 0.71 & 8.41 $\pm$ 4.07 \\
\quad EEGCT-Wide & 13.27 $\pm$ 2.16 & 8.27 $\pm$ 0.76 & 5.00 $\pm$ 2.22 \\
\quad RLSTM & 12.68 $\pm$ 2.30 & 8.38 $\pm$ 0.85 & 4.31 $\pm$ 2.38 \\
\quad STST & 14.95 $\pm$ 3.44 & 8.51 $\pm$ 0.93 & 6.44 $\pm$ 3.32 \\
\quad TSCNN & 12.68 $\pm$ 1.65 & 8.64 $\pm$ 0.45 & 4.04 $\pm$ 1.76 \\
\textbf{Additional models} & & & \\
\quad LR & 14.72 $\pm$ 2.07 & 8.34 $\pm$ 0.49 & 6.38 $\pm$ 2.35 \\
\quad kNN & 10.23 $\pm$ 0.92 & 8.41 $\pm$ 0.46 & 1.82 $\pm$ 1.14 \\
\quad SVC & 14.64 $\pm$ 2.14 & 8.55 $\pm$ 0.70 & 6.09 $\pm$ 2.31 \\
\quad ShallowConvNet & 11.82 $\pm$ 1.38 & 8.33 $\pm$ 0.60 & 3.49 $\pm$ 1.60 \\
\quad DeepConvNet & 9.92 $\pm$ 1.65 & 8.19 $\pm$ 0.69 & 1.72 $\pm$ 1.63 \\
\quad EEGNet & 14.19 $\pm$ 4.11 & 8.32 $\pm$ 0.47 & 5.88 $\pm$ 4.20 \\
\bottomrule
\end{tabular}

\end{table}

\section{Discussion}

While it can be readily observed from the results presented above that the mean decoding accuracies of all models were higher under the confounded cross-validation procedure, that this discrepancy is statistically-significant requires further analysis. In this section, we analyze and discuss the results of our experiments. However, as we observed discrepancies between the decoding accuracies obtained in our experiments and those reported in the affected publications, we first address the likely sources of these differences.

\subsection{Reconciling discrepancies with the affected studies}




While we designed a subset of our experiments to produce results which, like the affected studies, are affected by the repeated-stimulus confound, we observed discrepancies between the decoding accuracies of the models in these experiments, and those reported in the affected publications. For the LDA, ADCNN, AW1DCNN, and EEGCT models, the decoding accuracies we report in our experiments are within 4.5\% of the values reported in their corresponding publications. However, for the RLSTM, STST, and TSCNN models, the discrepancies in performance we observed were significantly larger, differing by 11\%--18\% from the reported values. These discrepancies are likely due to a combination of factors, including issues with reproducibility, differences in experimental design, the presence of additional confounds, and  publication bias.

While our first preference when replicating models from the affected works was the use of source code provided by the authors, we encountered substantial issues with the availability and completeness of source code. Of the models included in our experiments, source code was available for only the STST and TSCNN models. Moreover, we consider the source code in both cases to be incomplete as the implementations of the complex feature-engineering methods required to use these models were not provided. In the absence of source code, independent implementations were attempted using the descriptions provided in the corresponding publications. However, we found several key details omitted in both cases. For the remaining models, independent implementations of the model architectures were attempted. In the case of the RLSTM model, the provided description was not self consistent, and modifications were required to produce a working implementation. For the ADCNN model, the description of the mask applied to the data was ambiguous, and we cannot guarantee that the implementation we provide is identical to that used in the original study. The descriptions of the EEGCT and STST models were sufficient for unambiguous implementations, however, the feature-engineering methods were not described in sufficient detail to guarantee that the implementations we provide are identical to those used in the original study. While every care was taken to ensure that the implementations we provide are as close as possible to those used in the original studies, we cannot guarantee that they are identical, and these discrepancies may account for a portion of the differences in performance we observed.

Another potential factor contributing to the discrepancies we observed is the differences in experimental design between our experiments and those in the affected studies. To obtain robust estimates of the bias imparted by the RSC, we used a paired cross-validation procedure to train and evaluate models in all experiments. However, a consequence of this approach is that the confounded estimates of accuracy are obtained using an experimental design which differs in some respects from those used in the affected studies. However, the number of responses per stimulus used in training models, and obtaining confounded estimates of accuracy, is approximately the same as in the affected studies (64--65 vs.\ 66 responses per stimulus). Moreover, while we applied a consistent experimental design across all models, the affected publications report inconsistent model selection and evaluation methods, which are often affected by confounds other than the RSC. Consequently, we believe the most significant difference between the experimental design we report and those used in the affected works is the mitigation of these additional sources of bias.

In particular, we observed that many of the affected publications do not perform model selection using a validation set, but instead using the test set. This practice is referred to as \emph{post-hoc hyperparameter tuning}. The issue with this approach is that since the test set is used to select the model, models which overfit to the test set are more likely to be selected. And, as the number of hyperparameter combinations increases, the likelihood of a model substantially overfitting to the test set also increases. In the context of the RSC, this issue is even more severe. When the RSC is present, the test set shares stimulus-specific features with the training set, and therefore, the model most affected by the RSC is also the model most likely to be selected. Consequently, when post-hoc hyperparameter tuning is performed, it is likely that models are both less generalizable, and that their decoding accuracies are overestimated to a greater degree. All but two of the affected publications corresponding to the models in our experiments performed post-hoc hyperparameter tuning. One exception was the LDA model used in the original study. However, nested cross-validation was used to perform PCA in which the principal components were derived from the full dataset. Therefore, a separate form of test-set contamination is present in this case. The other exception was the RLSTM model; while hyperparameter values are reported, the authors do not specify how these values were selected.

Finally, a portion of the discrepancies we observed may be the result of publication bias arising from the extensive reuse of the SUD. Similarly to how models which overfit to a test set are more likely to be selected in post-hoc hyperparameter tuning, models which overfit to a dataset are more likely to be selected for publication. Consequently, over time, improvements in model performance on a benchmark dataset are less likely to indicate a genuine improvement in the ability of models to generalize to novel data. Publication bias is well documented in the case of the CIFAR-10 and ImageNet datasets, which were overused to the extent that model performance was overestimated by 3\%--15\% and 11\%--14\% respectively \cite{recht_imagenet_2019}. Moreover, when this is considered in conjunction with the RSC and the widespread use of post-hoc hyperparameter tuning in the affected studies, it seems likely that publication bias would favor models which overfit to stimulus-specific features of the SUD. If so, then the extent to which the RSC has affected the results reported in the affected studies would be greater than the results we report in our experiments.

\subsection{The repetition of stimuli across training and test sets biases estimates of category-decoding accuracy}

\begin{figure*}
  \centering
  \includegraphics[width=17.79cm, trim={0cm, 0cm, 0cm, 0cm}, clip]{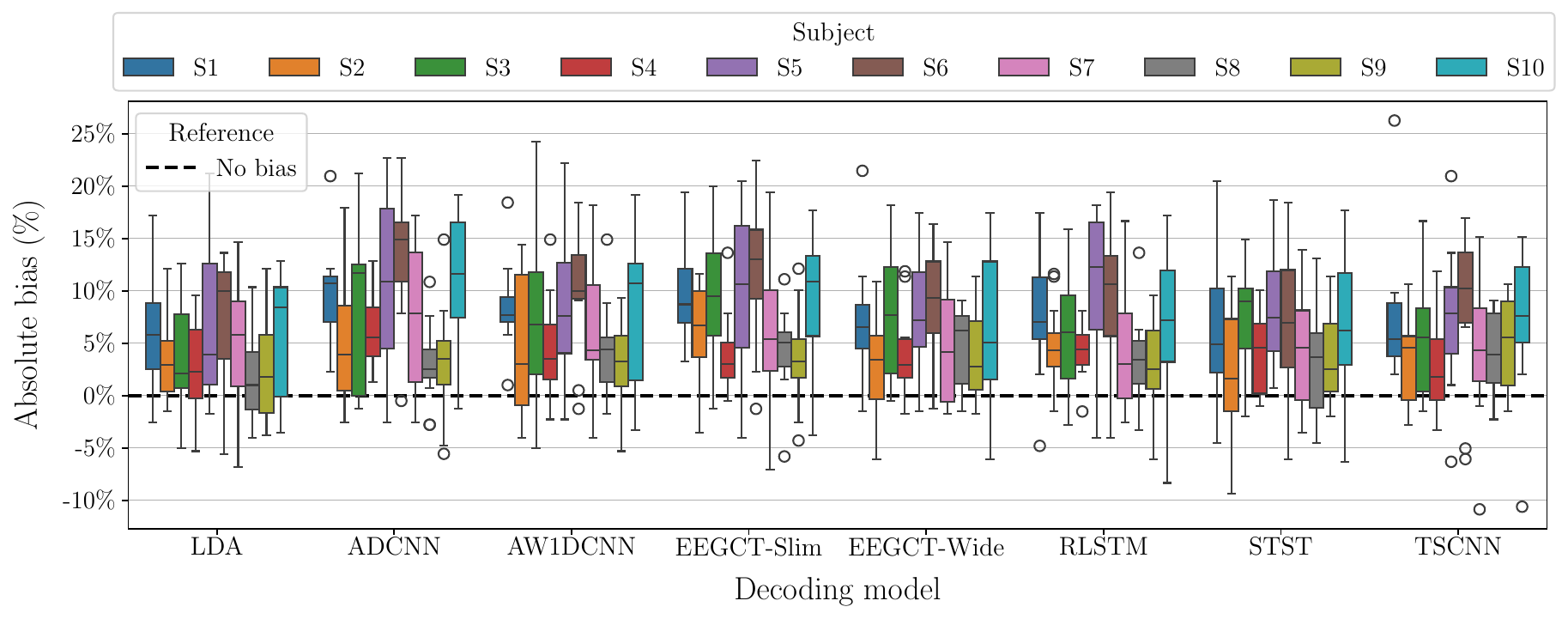}
  \caption{\textbf{Category-decoding accuracy bias by model and subject.} In our experiments, the bias in accuracy due to the confound is consistently positive across all models and subjects. We can conclude from this that the presence of bias is systematic, and does not depend on extraneous variables, such as model architecture or subject. Moreover, since the interquartile range of the bias for each subject rarely includes zero, this indicates that decoding accuracy is overestimated in over 75\% of all cross-validation folds. This consistency is even more significant than it may appear as, for each fold, the unconfounded estimate of accuracy is derived from responses to only one stimulus per category, and consequently is subject to a high degree of variability.}
  \label{fig:accuracy-bias-results}
\end{figure*}

Our primary claim in this work is that the repetition of stimuli across training and test sets results in biased estimates of category-decoding accuracy. We can observe from Fig.~\ref{fig:accuracy-bias-results} that, for all models, and over all subjects, the bias imparted by the confound is predominantly positive. This indicates that the repeated-stimulus confound has resulted in a consistent, and systematic, overestimation of decoding accuracy.

To support this claim, and to quantify the likely extent to which the decoding accuracies reported in the affected publications have been overestimated due to the confound, we performed hypothesis tests to determine if the bias affecting each model is significantly greater than zero. For each model, we applied a one-tailed t-test with confidence level $\alpha = 0.05$. To limit the impact of any variation in observed bias due to the stimuli used in the training and test sets of each fold, we averaged our estimates of bias across all folds, to obtain a single estimate of bias per subject. To account for the separate hypothesis tests performed for each model, we applied the Holm-Bonferroni procedure to adjust the significance levels of the tests~\cite{holm_simple_1979}. The results of the hypothesis tests are presented in Table~\ref{tab:bias-hypothesis-tests}. Additionally, we report the confidence intervals of the bias affecting each model using the Holm-Bonferroni adjusted significance levels.

\begin{table}
  \centering
  \caption{Results of hypothesis tests assessing the statistical significance of the bias affecting the estimated accuracy of each model.}
  \label{tab:bias-hypothesis-tests}
  \begin{tabular}{lcc}
\toprule
\textbf{Model} & \textbf{Bias estimate} & \textbf{95\% CI}\\
\midrule
\multicolumn{2}{l}{\textbf{Primary models}}\\
\quad LDA & 4.46\rlap{\textsuperscript{***}} & [2.08, 6.84]\\
\quad ADCNN & 7.94\rlap{\textsuperscript{***}} & [3.88, 12.00]\\
\quad AW1DCNN & 6.65\rlap{\textsuperscript{***}} & [4.90, 8.40]\\
\quad EEGCT-Slim & 7.42\rlap{\textsuperscript{***}} & [4.96, 9.88]\\
\quad EEGCT-Wide & 5.82\rlap{\textsuperscript{***}} & [3.23, 8.42]\\
\quad RLSTM & 6.04\rlap{\textsuperscript{***}} & [3.38, 8.71]\\
\quad STST & 5.17\rlap{\textsuperscript{***}} & [3.21, 7.13]\\
\quad TSCNN & 5.58\rlap{\textsuperscript{***}} & [3.37, 7.78]\\
\multicolumn{2}{l}{\textbf{Additional models}}\\
\quad LR & 5.72\rlap{\textsuperscript{***}} & [3.44, 7.99]\\
\quad kNN & 1.44\rlap{\textsuperscript{***}} & [0.66, 2.21]\\
\quad SVC & 5.42\rlap{\textsuperscript{***}} & [3.13, 7.70]\\
\quad ShallowConvNet & 2.99\rlap{\textsuperscript{***}} & [1.20, 4.78]\\
\quad DeepConvNet & 4.58\rlap{\textsuperscript{***}} & [2.67, 6.49]\\
\quad EEGNet & 6.32\rlap{\textsuperscript{***}} & [3.24, 9.40]\\
\bottomrule
\multicolumn{3}{p{0.7\linewidth}}{\small `***' indicates that the estimate is different from zero at the $p < 0.001$ significance level.}
\end{tabular}

\end{table}


The results of the hypothesis tests indicate that the bias in decoding accuracy imparted by the confound is significantly different from zero for all models. We can conclude that the repeated presentation of stimuli across training and test sets  has resulted in an overestimation of decoding accuracy for each of these models. Moreover, given the quantity and diversity of models used in our experiments, our results suggest that the presence of bias due to the confound is pervasive, and is likely to affect any decoding model evaluated using a confounded evaluation procedure. Furthermore, we can observe from our estimates of the bias affecting the models, and the corresponding confidence intervals, that the magnitude of the bias is considerable. For example, the bias affecting the ADCNN model is estimated to be 7.94\%, while our unconfounded estimate of its accuracy is 43.32\%. This implies that the confound resulted in a relative overestimation of accuracy of approximately 18.3\%.

\subsection{When the confound is present, the greater the estimated accuracy of a model, the more it has been overestimated}

\begin{figure*}
  \centering
  \includegraphics[width=17.79cm, trim={0cm, 0cm, 0cm, 0cm}, clip]{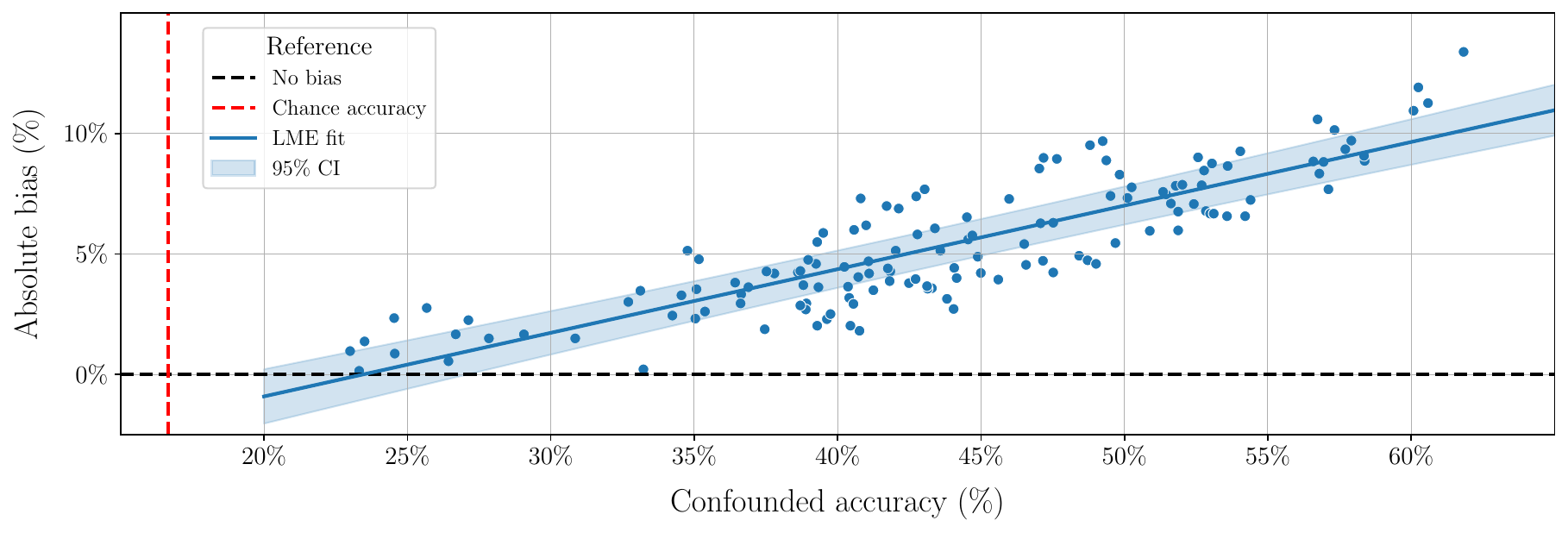}
  \caption{\textbf{Absolute bias vs.\ confounded accuracy.}
    When the confound is present, the greater the estimated accuracy of a model, the more it has been overestimated. This is indicated by the positive slope of a linear mixed-effects model fit to the data, shown above. The slope of the line indicates that, for every 1\% increase in confounded accuracy, the unconfounded accuracy is overestimated by approximately 0.26\%. The shaded 95\% confidence interval was calculated using critical values from the t-distribution ppf and the Satterthwaite degrees of freedom approximation.}
  \label{fig:bias-vs-confounded-accuracy}
\end{figure*}

We also asked if models which achieved higher estimates of decoding accuracy under the confounded cross-validation procedure were biased to a greater extent than less performant models. We considered this an important question to address, as this would indicate that the accuracies of state-of-the-art models are overestimated to a greater extent as new, more proficient models are developed. This, in combination with the potential for publication bias, would suggest that reported improvements in decoding accuracies are less indicative of true improvements in model performance over time. And, as Fig.~\ref{fig:bias-vs-confounded-accuracy} illustrates, the confounded estimates of decoding accuracy do appear to increase at a greater rate than the unconfounded estimates.


To quantify the relationship between the confounded and unconfounded estimates of decoding accuracy, we performed a linear-regression analysis. We used a linear mixed-effects model which included random effects for the model and subject variables. As in the previous analysis, we averaged our estimates across all folds, to obtain a single estimate of each variable per subject and eliminate the impact of any variation due to the stimuli used in the training and test sets of each fold. We also included a fixed effect for the intercept term, as we expected that the bias would be non-zero even when the confounded estimate of accuracy is zero. We did not include a random effect for the fold variable, as we expected that the magnitude of the bias would be independent of the fold. This is because we expect that the bias imparted by the confound is due to the model architecture and subject, rather than the specific model instance trained on each fold.

In this way, we account for the variance due to the model architecture and subject, as well as the variance due to the specific model instance trained on each fold. This model is described by the equation
\begin{equation}
  \text{Z}_{i, j} = \beta_0 + \beta_1 \text{X}_{i, j} + m_i + s_j +\epsilon_{i, j}, \label{eqn:accuracy_eqn}
\end{equation}
where, for the $j$th subject for the $i$th model, $\text{X}_{i, j, k}$ and $\text{Z}_{i, j}$ are the accuracy under the confound and bias respectively. The intercept is given by $\beta_0$, and $\beta_1$ is the fixed effect for a unit increase in accuracy under the confound. The random effects for the model and subject variables are given by $m_i\sim\mathcal{N}(0, \sigma_{m}^2)$ and $s_{j}\sim\mathcal{N}(0, \sigma_{s}^2)$, respectively, and the residual error is given by $\epsilon_{i, j}\sim\mathcal{N}(0, \sigma^2)$. Our null hypothesis was that $\beta_1 = 0$, which implies that the confounded estimate of accuracy is directly proportional to the unconfounded estimate. To fit this model, as well as subsequent linear mixed-effects models, we used the R package \texttt{lme4} \cite{bates_fitting_2015}. Additionally, 95\% confidence intervals were constructed for the fixed-effects using the Wald method~\cite{wald_tests_1943}. The results of the linear mixed-effects model are presented in Table~\ref{tab:bias-vs-confounded-lme}.

\begin{table}
  \centering
  \caption{Results of the linear mixed-effects regression analysis assessing the change in bias relative to accuracy under the confound.}
  \label{tab:bias-vs-confounded-lme}
  \begin{tabular}{lr@{\hspace{3em}}c}
\toprule
\textbf{Parameter} &
\multicolumn{1}{l}{\textbf{Estimate}} &
\textbf{95\% CI} \\
\midrule
\textbf{Fixed effects} & & \\
\quad $\beta_{0}$ & $-$6.19 & [$-$7.83, $-$4.55] \\
\quad $\beta_{1}$ & 0.26\rlap{\textsuperscript{***}} & [0.23, 0.30] \\
\textbf{Random effects} & & \\
\quad $\sigma^2_{m}$ & 0.21 & \\
\quad $\sigma^2_{s}$ & 1.21 & \\
\quad $\sigma^2$ & 0.59 & \\
\bottomrule
\multicolumn{3}{p{0.7\linewidth}}{\small `***' indicates that the estimate is different from zero at the $p < 0.001$ significance level.}
\end{tabular}

\end{table}


The results of our analysis indicate that, under the confound, the magnitude of bias increases significantly with accuracy. Our estimate of the fixed effect $\beta_1$ suggests that, under the confound, given two models with a 1\% difference in decoding accuracy, the actual difference in their performance is approximately 0.76\%. Equivalently, we can conclude that improvements in model performance are overestimated by a factor of approximately 1.35 under the confound.


As mentioned previously, we were unable to reproduce the high estimates of decoding accuracy reported in several of the affected publications, likely due to reproducibility issues, or additional sources of bias. However, the results of our analysis suggest that, assuming the results reported in the literature were reproducible, and otherwise valid, they would be overestimated by a greater margin than we observed in our experiments. For example, the STST model was reported to achieve a decoding accuracy of 54.82\%, and so we would infer that this value is overestimated by approximately 8.06\%, and the actual accuracy of the model is approximately 46.76\%.

\subsection{The relative difficulty of decoding different object categories is underestimated under the confound}

\begin{figure*}
  \centering
  \includegraphics[width=11.4cm, trim={0cm, 0cm, 0cm, 0cm}, clip]{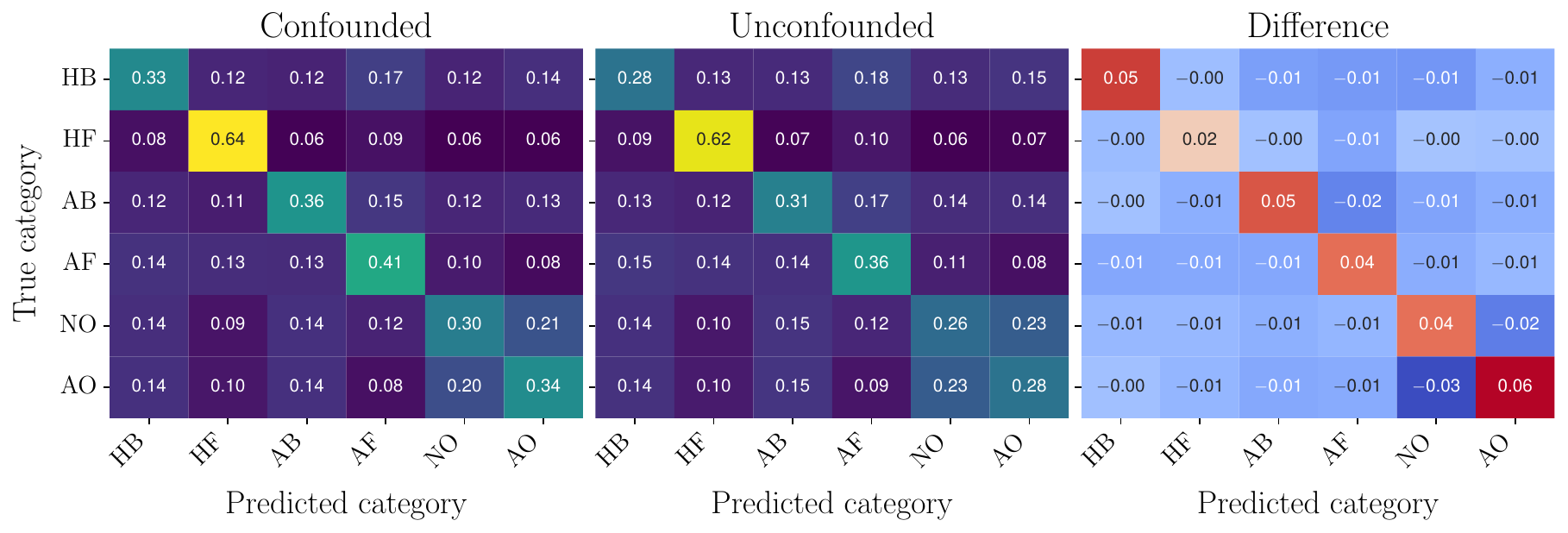}
  \caption{\textbf{Confounded vs.\ unconfounded confusion matrices for the LDA model.}
    Under the confound (left), the Artificial Object and Natural Object categories appear relatively separable, with both classified correctly more often than they were misclassified as each other. However, in the absence of the confound (middle), they are almost misclassified as each other as often as they are correctly classified. The difference between the two confusion matrices (right) clearly shows that each category achieves a lower classification rate when the confound is not present.}
  \label{fig:confounded-vs-unconfounded-confusion-matrix}
\end{figure*}

We also asked if the magnitude of the bias due to the confound was dependent on object category. We considered this an important question to address; if the bias imparted by the confound differed between object categories then not only would the performance of the models in the affected publications have been overestimated, but any claims regarding the relative difficulty of the different object categories would be called into question. This is particularly relevant to the original study for which the SUD was created, as the authors of this work use the misclassification rates of different object categories to assess the separability and similarity of the different object categories. And, the confounded and unconfounded confusion matrices for the LDA model presented in Fig.~\ref{fig:confounded-vs-unconfounded-confusion-matrix} indicate the bias imparted by the confound does vary by object category. Interestingly, while investigating the implications of the RSC for the original study, we discovered that the reported measure of dissimilarity is invalid. As a consequence, the representational dissimilarity matrices constructed using this method feature negative self dissimilarity for several categories, and incomparable measures of dissimilarity between categories.

To confirm the significance of the dependence of bias on object category across all models, we fit a linear mixed-effects model. As in the previous analysis, we included random effects for the model and subject variables. The model is described by the equation
\begin{equation}
  \text{Z}_{i, j, k} = \beta_0 + \sum_{c=1}^{5}\beta_{c} \cdot C_{k} + m_{i} + s_{j} +\epsilon_{i, j, k}, \label{eqn:category-bias_eqn}
\end{equation}
where $\text{Z}_{i, j, k}$ is the estimated bias for the $k$th category of the $j$th subject for the $i$th model. Here $\beta_0$ is the fixed effect for the reference category, and $\beta_{c}$ is the fixed effect for the $c$th category, relative to the reference category. We use $C$ as a dummy variable for object category, such that $C_{x}$ is equal to 1 if $x=K$, and 0 otherwise. The random effects for the model and subject variables are given by $m_i\sim\mathcal{N}(0, \sigma_{m}^2)$, and $s_i\sim\mathcal{N}(0, \sigma_{s}^2)$, respectively. And, the residual error is then given by $\epsilon_{i, j, k}\sim\mathcal{N}(0, \sigma^2)$. The results of the linear mixed-effects model are presented in Table~\ref{tab:category-bias-lme}.


After fitting the linear mixed-effects model, we applied the Holm-Bonferroni post-hoc method to assess if the difference in bias between any two pairs of categories was significant. This revealed that the magnitude of the bias for the Human Face category is significantly lower than for all other categories at the $p < 0.001$ significance level. Moreover, as most pairs of categories exhibit statistically-significant differences in bias, the hypotheses tests establish that the bias imparted by the confound is highly dependent on object category. The results of the analysis are presented in Table~\ref{tab:category-bias-lme}. Additionally, the 95\% confidence intervals we report for the fixed-effects were constructed using the Wald method~\cite{wald_tests_1943}. However, the confidence intervals for the contrast between fixed-effects were constructed using the Holm-Bonferroni adjusted significance levels used to perform the hypothesis tests.

\begin{table}
  \centering
  \caption{Results of the linear mixed-effects regression analysis assessing the dependence of the bias imparted by the confound on object category.}
  \label{tab:category-bias-lme}
  \begin{tabular}{lrc}
\toprule
\textbf{Parameter} & \textbf{Estimate} & \textbf{95\% CI}\\
\midrule
\textbf{Fixed effects} & & \\
\quad Animal Body & 5.94 & [4.28, 7.61] \\
\quad Animal Face & 4.55 & [2.89, 6.21] \\
\quad Artificial Object & 7.65 & [5.99, 9.31] \\
\quad Human Body & 6.44 & [4.78, 8.10] \\
\quad Human Face & 1.98 & [0.31, 3.64] \\
\quad Natural Object & 5.81 & [4.15, 7.48] \\
\textbf{Contrast} & & \\
\quad Animal Body---Animal Face & 1.39\rlap{\textsuperscript{***}} & [0.57, 2.21] \\
\quad Animal Body---Artificial Object & $-$1.71\rlap{\textsuperscript{***}} & [$-$2.53, $-$0.89] \\
\quad Animal Body---Human Body & 0.50\rlap{\textsuperscript{}} & [$-$0.32, 1.32] \\
\quad Animal Body---Human Face & $-$3.97\rlap{\textsuperscript{***}} & [$-$4.79, $-$3.15] \\
\quad Animal Body---Natural Object & 0.13\rlap{\textsuperscript{}} & [$-$0.69, 0.95] \\
\quad Animal Face---Artificial Object & $-$3.10\rlap{\textsuperscript{***}} & [$-$3.92, $-$2.28] \\
\quad Animal Face---Human Body & 1.89\rlap{\textsuperscript{***}} & [1.07, 2.71] \\
\quad Animal Face---Human Face & $-$2.58\rlap{\textsuperscript{***}} & [$-$3.40, $-$1.75] \\
\quad Animal Face---Natural Object & $-$1.26\rlap{\textsuperscript{***}} & [$-$2.08, $-$0.44] \\
\quad Artificial Object---Human Body & $-$1.21\rlap{\textsuperscript{***}} & [$-$2.03, $-$0.39] \\
\quad Artificial Object---Human Face & $-$5.67\rlap{\textsuperscript{***}} & [$-$6.50, $-$4.85] \\
\quad Artificial Object---Natural Object & $-$1.84\rlap{\textsuperscript{***}} & [$-$2.66, $-$1.02] \\
\quad Human Body---Human Face & 4.46\rlap{\textsuperscript{***}} & [3.65, 5.29] \\
\quad Human Body---Natural Object & 0.63\rlap{\textsuperscript{.}} & [$-$0.19, 1.45] \\
\quad Human Face---Natural Object & $-$3.84\rlap{\textsuperscript{***}} & [$-$4.66, $-$3.02] \\
\textbf{Random effects} & &\\
\quad $\sigma_{s}^2$ & 2.74 & \\
\quad $\sigma_{m}^2$ & 3.95 & \\
\quad $\sigma^2$ & 5.44 & \\
\bottomrule
\multicolumn{3}{p{0.85\linewidth}}{\small `***' indicates that the estimate is different from zero at the $p < 0.001$ significance level.}
\end{tabular}

\end{table}

\begin{figure*}
  \centering
  \includegraphics[width=11.4cm, trim={0cm, 0cm, 0cm, 0cm}, clip]{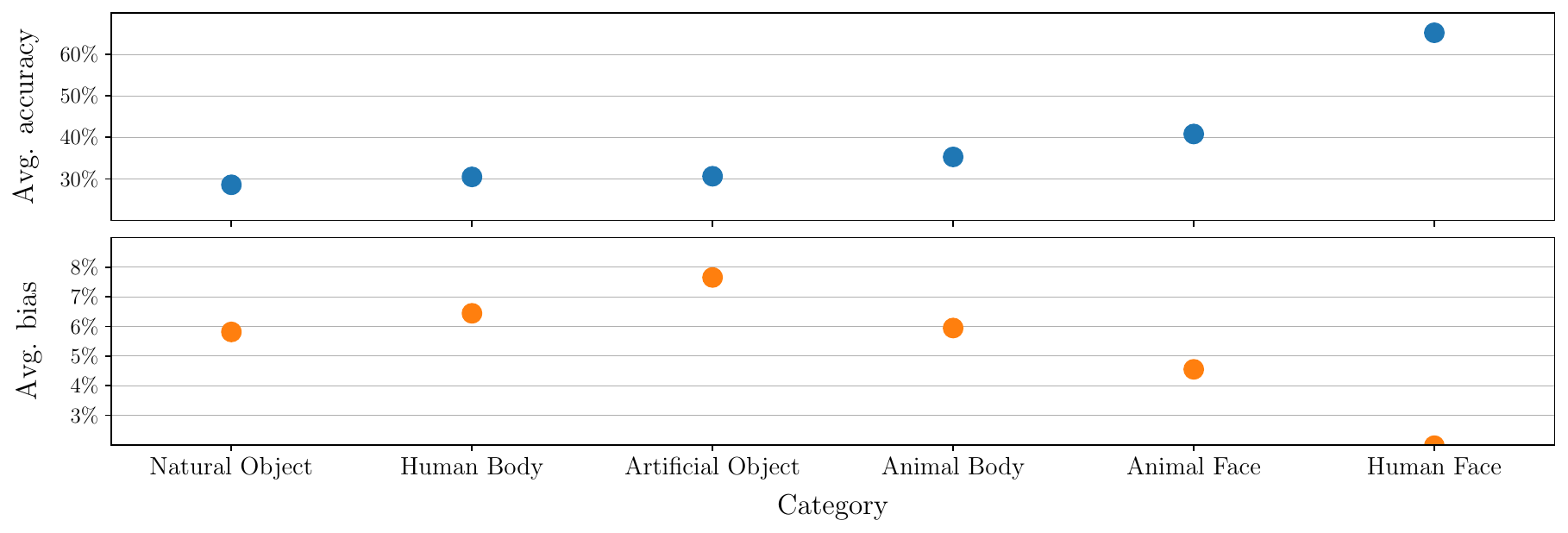}
  \caption{\textbf{Mean unconfounded accuracy and bias over all models by object category.} When the confound is absent, the less often models were able to correctly decode object categories, the more the accuracy on that category tended to be overestimated when the confound was present. This indicates that the difficulty of decoding the hardest object categories has been substantially underestimated in the literature.}
  \label{fig:category-bias-accuracy}
\end{figure*}

Interestingly, while our prior analysis demonstrated that, when considering decoding accuracy over all categories, the most proficient models are those most affected by the confound, the more easily decodable a category is, the less it tends to be affected by the confound. This is illustrated in Fig.~\ref{fig:category-bias-accuracy}, which shows the mean unconfounded accuracy and bias for each object category across all models. This may result in the appearance of a more uniform ability to decode different object categories than is actually the case. A consequence of this is that the difficulty of decoding the hardest object categories has been underestimated to a greater extent than our prior analyses would suggest.

As mentioned previously, this is of particular relevance to the original study for which the SUD was collected, as the primary analyses presented in the publication rely on performing RSA using the misclassification rate between categories as a measure of similarity. Consequently, as the bias imparted by the confound is dependent on object category, the results of these studies are also biased. Interestingly, while investigating the extent of this bias, we discovered that the measure of dissimilarity used in the original work is invalid. As a consequence, representational dissimilarity matrices constructed using this method feature negative self dissimilarity for several categories, and incomparable measures of dissimilarity between categories.

\subsection{Confounded estimates of decoding accuracy could be used to support claims of extrasensory perception}

The publications which we investigate in this study all perform object-category decoding, and our experiments, as well as prior results in the literature, indicate that EEG recordings of responses to object categories are, to some extent, separable \cite{daliri_eeg_2013}. However, the repeated-stimulus confound is not restricted solely to studies which perform object-category decoding. The confound may affect the results of any decoding experiment in which the stimuli are repeated across training and test sets. Consequently, we asked what the implications of the confound would be for studies which perform novel decoding tasks, which are attempting to decode classes of unknown separability. In particular, could the results of an analysis confounded by the repetition of stimuli be used to support pseudoscientific claims, such as the existence of extrasensory perception?

For example, suppose a study wished to investigate if humans can subconsciously sense the astrological signs of others. If the study were to present images of human faces, and then attempt to decode the astrological sign of the person depicted in the image, would the repetition of the same faces across training and test sets result in a significant decoding accuracy? If so, a study, with potentially more scientific rigor than many of the affected publications, could use these results to support this pseudoscientific claim.

This would be a serious issue as the work could even reference the methodology of the affected publications as a precedent for the validity of their analysis. Consequently, if an article describing the study were to be submitted for publication at a venue which published one of the affected studies, it would be difficult to argue against the validity of its claims.

Therefore, to determine if results obtained under the confound could support claims that non-existent properties of stimuli are decodable, we designed a novel, and impossible, decoding task. We refer to this task pseudocategory decoding. Similarly to the category-decoding task, the pseudocategory-decoding task involves training a model to predict the category of a stimulus. However, by design, the pseudocategories are meaningless, and generalization to novel stimuli is impossible. In our pseudocategory-decoding experiments, we found that all models achieved a decoding accuracy greater than chance level (8.25\%) when the confound was present.

To assess the significance of our results, we performed hypothesis tests to assess if the decoding accuracy of each model was significantly different from chance level. We performed a separate test for the confounded and unconfounded results of each model. For each model, we performed a one-tailed t-test with an initial confidence level of $\alpha = 0.05$, and adjusted the significance level for the tests using the Bonferroni method. Additionally, the 95\% confidence intervals for the decoding accuracies of each model were constructed using the Bonferroni-corrected significance level. The results of the hypothesis tests are presented in Table~\ref{tab:pseudocategory-hypothesis-tests}.

\begin{table}
  \centering
  \caption{Results of hypothesis tests assessing if each model could decode pseudocategory identity with above chance accuracy under confounded and unconfounded evaluation procedures.}
  \label{tab:pseudocategory-hypothesis-tests}
  \begin{tabular}{l@{\hspace{0em}}r@{\hspace{3em}}c@{\hspace{0em}}r@{\hspace{2em}}c}
\toprule
& \multicolumn{2}{c}{\textbf{Confounded}} & \multicolumn{2}{c}{\textbf{Unconfounded}}\\
\cmidrule(lr){2-3} \cmidrule(lr){4-5}
\textbf{Model} & \multicolumn{1}{l}{\textbf{Accuracy}} & \textbf{95\% CI} & \multicolumn{1}{l}{\textbf{Accuracy}} & \textbf{95\% CI}\\
\midrule
\textbf{Affected models}\\
\quad LDA & 10.90\rlap{\textsuperscript{*}} & [8.57, 13.22] & 8.79\rlap{\textsuperscript{}} & [7.97, 9.60] \\
\quad ADCNN & 19.50\rlap{\textsuperscript{**}} & [12.88, 26.13] & 8.49\rlap{\textsuperscript{}} & [7.40, 9.58] \\
\quad AW1DCNN & 15.32\rlap{\textsuperscript{**}} & [10.77, 19.88] & 8.40\rlap{\textsuperscript{}} & [7.44, 9.35] \\
\quad EEGCT-Slim & 16.58\rlap{\textsuperscript{**}} & [11.26, 21.90] & 8.17\rlap{\textsuperscript{}} & [7.19, 9.15] \\
\quad EEGCT-Wide & 13.27\rlap{\textsuperscript{**}} & [10.28, 16.25] & 8.27\rlap{\textsuperscript{}} & [7.21, 9.32] \\
\quad RLSTM & 12.68\rlap{\textsuperscript{**}} & [9.50, 15.86] & 8.38\rlap{\textsuperscript{}} & [7.20, 9.55] \\
\quad STST & 14.95\rlap{\textsuperscript{**}} & [10.19, 19.70] & 8.51\rlap{\textsuperscript{}} & [7.22, 9.79] \\
\quad TSCNN & 12.68\rlap{\textsuperscript{***}} & [10.40, 14.96] & 8.64\rlap{\textsuperscript{}} & [8.01, 9.27] \\
\textbf{Additional models} & & & & \\
\quad LR & 14.72\rlap{\textsuperscript{***}} & [11.86, 17.57] & 8.34\rlap{\textsuperscript{}} & [7.66, 9.02] \\
\quad kNN & 10.23\rlap{\textsuperscript{**}} & [8.95, 11.51] & 8.41\rlap{\textsuperscript{}} & [7.78, 9.04] \\
\quad SVC & 14.64\rlap{\textsuperscript{***}} & [11.69, 17.59] & 8.55\rlap{\textsuperscript{}} & [7.57, 9.52] \\
\quad ShallowConvNet & 11.82\rlap{\textsuperscript{***}} & [9.91, 13.73] & 8.33\rlap{\textsuperscript{}} & [7.50, 9.17] \\
\quad DeepConvNet & 9.92\rlap{\textsuperscript{}} & [7.63, 12.20] & 8.19\rlap{\textsuperscript{}} & [7.24, 9.15] \\
\quad EEGNet & 14.19\rlap{\textsuperscript{*}} & [8.51, 19.88] & 8.32\rlap{\textsuperscript{}} & [7.67, 8.96] \\
\bottomrule
\multicolumn{5}{p{0.9\linewidth}}{\small `*', `**', and `***' indicate that the estimate is different from chance level (8.25\%) at the $p < 0.05$, $p < 0.01$, and $p < 0.001$ significance levels, respectively.}
\end{tabular}

\end{table}


Our results indicate that, when the confound was present, the decoding accuracies of all models, except for DeepConvNet, are significantly greater than chance level at the $p < 0.05$ significance level. However, under the unconfounded procedure, the decoding accuracy of all models is not significantly greater than chance level at the $p < 0.05$ significance level. While the results obtained under the confounded procedure do not directly support claims of extrasensory perception, they do indicate that stimulus-specific features of the training set can be learned by a model, even when there is no underlying structure to the class labels. Consequently, we can conclude, at least in the case of images of natural objects, that results obtained when the repeated-stimulus confound provide evidence that non-existent properties of stimuli are decodable. Therefore, the confound could be used to support an array of pseudoscientific claims, including the existence of extrasensory perception. And, as expected, we can conclude that the results obtained using our solution to the confound do not support such claims.

\subsection{Future work}

As a result of our investigation, we have identified three key areas which require further investigation.

Firstly, the repeated-stimulus confound is a pervasive issue in the field of decoding experiments, and is likely to affect the results in studies other than those which use SUD. Moreover, while the repetition of stimuli was the focus of our investigation, it is possible that, even when stimuli are not repeated, the repetition of stimulus properties may result in a similar confound. This is of particular relevance, as increasingly large stimulus sets are being used to collect EEG data. And, as the number of stimuli increases, the more likely the repetition of some stimulus properties is. And, even if these properties are not repeated across training and test sets, models may still learn to overfit to them.

For example, the THINGS-EEG dataset \cite{herbart_things_2019} is a large open-access dataset which contains EEG recordings of responses to 1,854 concepts, each represented by 12 unique images. And, given the hierarchical nature of the stimulus set, while more abstract concepts are represented by a greater number of unique images, they are composed of a lesser quantity of more concrete concepts which may share confounding features. Therefore, while one of stated potential uses of the dataset is to develop models for encoding semantic information, the repetition of concepts presents significant challenges in this regard. Consequently, an important avenue for future work is the development of a robust set of guidelines for validating the reliability of encoding and decoding models.

Secondly, while our paired cross-validation procedure is effective in quantifying the stimulus-specific bias of a decoding model, we were unable to successfully extend this solution to prevent models from overfitting to stimulus-specific features. This is of most significance when training encoding or decoding models for use in applications like model-based RSA. In these contexts, since the goal is to develop a model which learns generalizable embeddings of the target variable, the presence of stimulus-specific bias is a threat to the validity of the learned representations. Therefore, an important avenue for future work is the development of a regularization procedure which penalizes stimulus-specific bias.

Lastly, the issue of post-hoc hyperparameter tuning affects many of the publications we investigated, and is likely a widespread issue within the literature. Therefore, a broader review of the literature on decoding studies to determine the prevalence of post-hoc hyperparameter tuning, and the extent to which it has biased the results of affected publications, is an important direction for future research. In particular, we are interested in investigating if post-hoc hyperparameter tuning exacerbates the impact of confounds in vulnerable datasets such as the SUD.

\section{Conclusion}

In this work, we investigated the repeated-stimulus confound, a pervasive issue which has likely led to a substantial overestimation of decoding accuracy reported in several publications. We also demonstrated that when the confound is present, the more performant a model appears to be, the more its performance has been overestimated. Similarly, our analysis revealed that the relative difficulty of decoding different object categories has been substantially underestimated. Consequently, many of the claims made in the affected publications are of questionable validity. Moreover, we demonstrated that the confounded methodology used in the affected studies could even be used to support pseudoscientific claims. While our audit of the literature was restricted to a single susceptible dataset, we determined that at least 18 publications are affected by the confound. However, we also identify at least three other popular datasets which are also susceptible to the confound, and so it is likely that the issue is significantly more widespread than our preliminary investigation suggests. Therefore, given the current extent to which the confound has affected results in the literature, and the potential severity of future misuse of the confounded methodology, it is important that our findings are disseminated to caution the wider community. The urgency of this is underscored by the issues we encountered when attempting to reproduce the results of these publications.

To facilitate independent re-evaluation of work which utilizes susceptible datasets, we presented a method for mitigating the confound which can be easily implemented using existing software packages. We also demonstrated that this method can be extended to validate the stimulus-specific bias of decoding models. Finally, we discussed the possibility that the repeated-stimulus confound may be a specific example of a more general confound which may affect a broader class of decoding studies. Thus, we consider our investigation an important step in a more large-scale investigation of the reproducibility and reliability of publications in systems neuroscience.

\section*{Acknowledgments}

We would like to thank Hari Bharadwaj for suggesting we investigate the SUD due to its small stimulus set and high ratio of repetitions to unique stimuli. This publication has emanated from research supported in part by research grants from Taighde Éireann --- Research Ireland under Grant Numbers 16/RI/3399, 18/CRT/6049, and 20/FFP-P/8853, the US National Science Foundation under Grant Number 1734938-IIS, and the Defense Advance Research Projects Agency under Grant Number 13001129 (prime contract award HR0011222003, subcontract award 2103299-01). The views, opinions, and/or findings expressed are those of the authors and should not be construed as a position or endorsement of any funding agency or government.

\bibliographystyle{abbrvnat}
\bibliography{tpami2025}

\newpage

\vspace{11pt}

\begin{IEEEbiography}
  [{\includegraphics[width=1in,height=1.25in,clip,keepaspectratio]
      {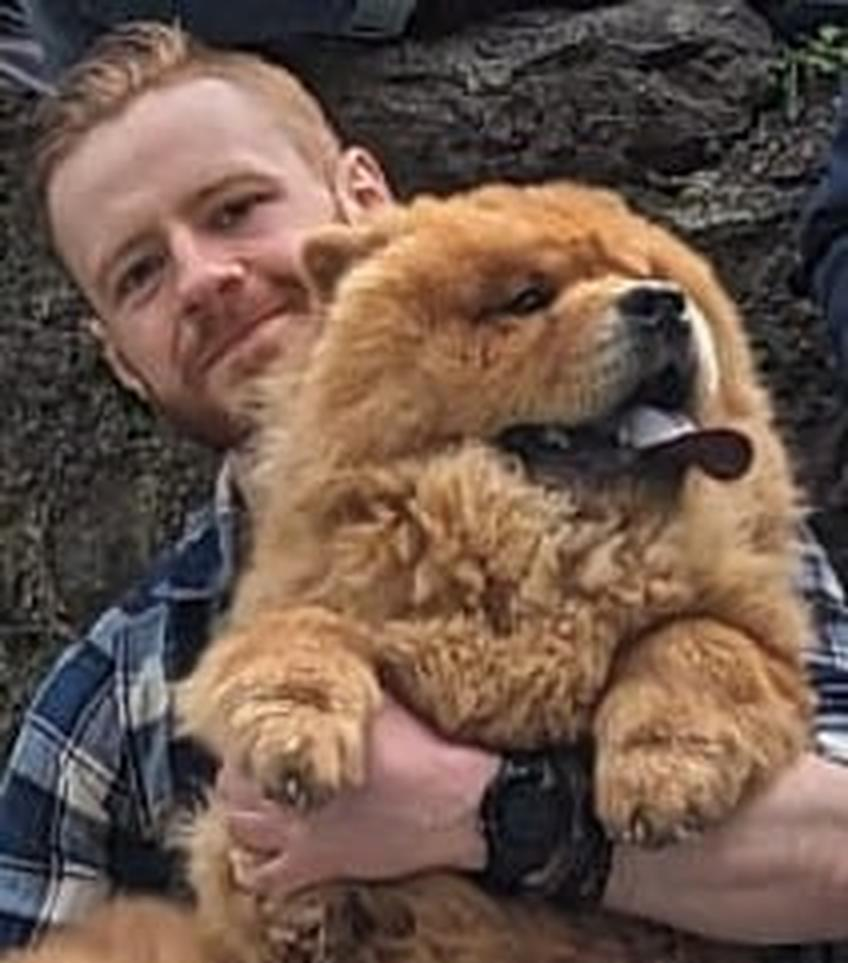}}]{Jack A. Kilgallen} received a BSc in Computational Thinking from Maynooth University in Ireland in 2021.
      He is currently a PhD candidate in the Hamilton Institute, also at Maynooth University, where he is investigating the impact of confounds on the reliability of decoding models in systems neuroscience.
\end{IEEEbiography}

\begin{IEEEbiography}
  [{\includegraphics[width=1in,height=1.25in,clip,keepaspectratio]
      {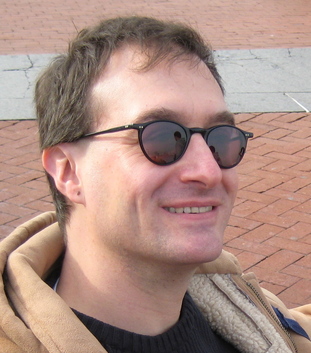}}]{Barak A.\ Pearlmutter} received a BS in Mathematics from Case Western Reserve University, a PhD in Computer Science from Carnegie Mellon University (where he worked on neural networks the second time they were cool), postdoctoral training in Neuroscience at Yale University, and spent several years in Industry at Siemens Corp Research and on the faculty in the Dept of Computer Science at the University of New Mexico.  He is currently a professor in the Dept of Computer Science at Maynooth University in Ireland.
\end{IEEEbiography}

\begin{IEEEbiography}
  [{\includegraphics[width=1in,height=1.25in,clip,keepaspectratio]
      {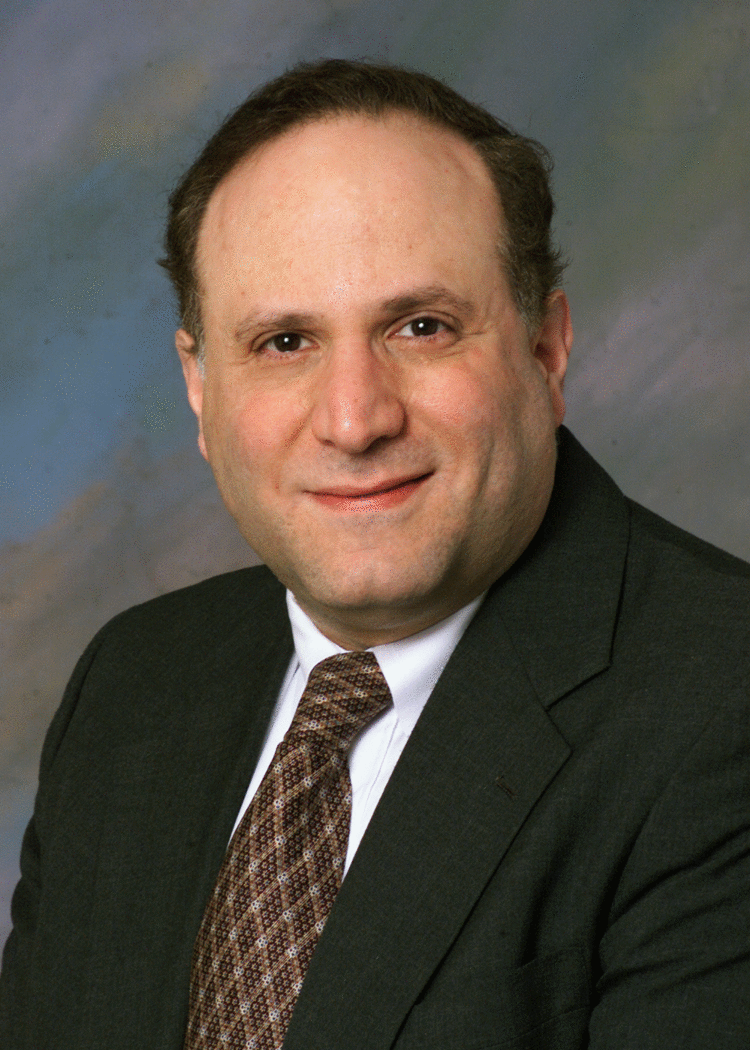}}]{Jeffrey Mark Siskind} received the B.A. degree in
  Computer Science from the Technion, Israel Institute of Technology in 1979,
  the S.M.\ degree in Computer Science from MIT in 1989, and the Ph.D.\ degree
  in Computer Science from MIT in 1992.
  He did a postdoctoral fellowship at the University of Pennsylvania
  Institute for Research in Cognitive Science from 1992 to 1993.
  He was an assistant professor at the University of Toronto Department of
  Computer Science from 1993 to 1995, a senior lecturer at the Technion
  Department of Electrical Engineering in 1996, a visiting assistant professor
  at the University of Vermont Department of Computer Science and Electrical
  Engineering from 1996 to 1997, and a research scientist at NEC Research
  Institute, Inc.\ from 1997 to 2001.
  He joined the Purdue University School of Electrical and Computer
  Engineering in 2002 where he is currently a professor.
\end{IEEEbiography}

\vfill

\end{document}